\documentclass[a4paper]{jpconf} 
\usepackage{graphicx}
\usepackage{amsmath}
\usepackage{hyperref}
\usepackage{amssymb}
\usepackage{wasysym}
\usepackage{gensymb}

\begin{document}

\title{The BINGO telescope: a new instrument exploring the new $21~cm$ cosmology window}

\author{C.A. Wuensche and the BINGO collaboration\footnote{http://www.bingotelescope.org}}

\address{Instituto Nacional de Pesquisas Espaciais, Divisão de Astrofísica. Av. dos Astronautas 1758, Jd. da Granja, CEP 12227-010, São José dos Campos, SP, Brazil}

\ead{ca.wuensche@inpe.br}

\begin{abstract}
BINGO is a unique radio telescope designed to make the first detection of Baryon Acoustic Oscillations (BAO) at radio frequencies. This will be achieved by measuring the distribution of neutral hydrogen gas at cosmological distances using a technique called Intensity Mapping. Along with the Cosmic Microwave Background anisotropies, the scale of BAO is one of the most powerful probes of cosmological parameters, including dark energy. The telescope will be built in a very low RFI site in South America and will operate in the frequency range from 0.96 GHz to 1.26 GHz. The telescope design consists of  two $\thicksim$ 40-m compact mirrors with no moving parts. Such a design will give the excellent polarization performance and very low sidelobe levels required for intensity mapping. With a feedhorn array of 50 receivers, it will map a $15^{\circ}$ declination strip as the sky drifts past the field-of-view of the telescope. The BINGO consortium is composed Universidade de S\~ao Paulo, Instituto Nacional de Pesquisas Espaciais (Brazil), University of Manchester and University College London (United Kingdom), ETH Z\"urich (Switzerland) and Universidad de La Republica (Uruguay). The telescope assembly and horn design and fabrication are under way in Brazil. The receiver was designed in UK and will be developed in Brazil, with most of the components for the receiver will also be supplied by Brazilian industry. The experience and science goals achieved by the BINGO team will be advantageous as a pathfinder mission for the Square Kilometre Array (SKA) project. This paper reports the current status of the BINGO mission, as well as preliminary results already obtained for the instrumentation development. 
\end{abstract}

\section{Introduction}
One of the main cosmological challenges in the 21st century is the explanation of the cosmic acceleration, first unequivocally inferred in 1998 by two independent groups measuring supernovae of type Ia \cite{Perlmutter1998, Riess1998}. This led to the 2011 Nobel Prize in Physics and can be explained by postulating a negative pressure from a new component, known as dark energy. In combination with other observations, such as the Cosmic Microwave Background (CMB), there is little doubt about the existence of such a component and the main focus of observational cosmology is now to try to determine its detailed properties. 

Among the various programs to measure those properties, the study of Baryonic Acoustic Oscillations (BAO) is recognized as one of the most powerful probes of the properties of dark energy (e.g. \cite{Albrecht2006}).To date, BAOs have only been detected by performing large galaxy redshift surveys in the optical waveband and it is important they are confirmed in other wavebands and measured over a wide range of redshifts. The radio band provides a unique and complementary observational window for the understanding of dark energy via the redshifted $21~cm$ neutral hydrogen emission line from distant galaxies. Most important, no detection of BAOs in radio has been claimed so far. 

\begin{figure}[h]
\begin{center}
\includegraphics[width=10cm]{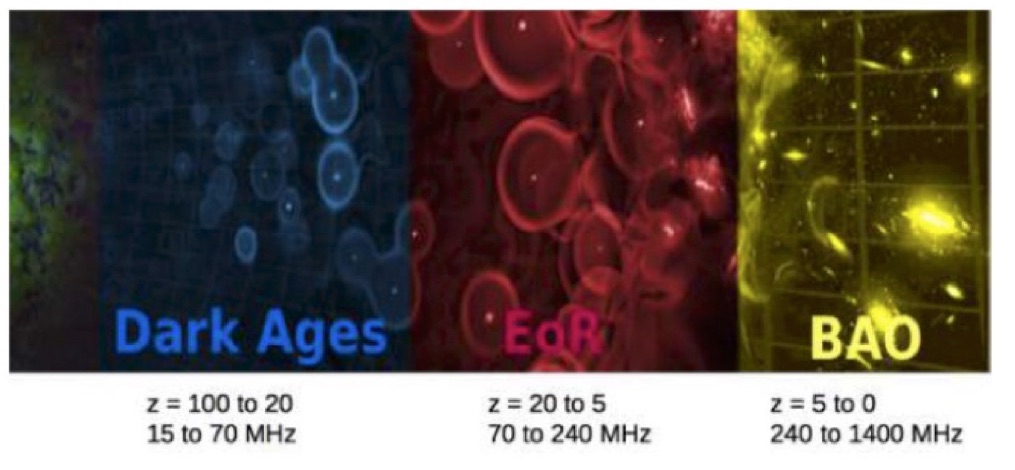}
\caption{Late history of the Universe and the radio frequency of observation of the 21cm line. The BAO epoch, well after reionization, will be the target of the BINGO telescope. In this epoch dark energy is dominating the expansion of the universe and it is possible to probe it.}
\label{BAOhistory}
\end{center}
\end{figure}

BAOs are a signature in the matter distribution from the recombination epoch (see Fig. \ref{BAOhistory}). Note that BAO show up in the distribution of galaxies at redshifts less than that of the epoch of reionization ($z \lesssim 5$). In the context of the standard cosmological model, the so-called $\Lambda-\textrm{CDM}$ model, BAOs manifest themselves as a small but detectable excess of galaxies with separations of order of $150~\textrm{Mpc.h}^{-1}$ This excess is the imprint of the acoustic oscillations generated during CMB times and its linear scale is known from basic physics. Consequently, a measure of its angular scale can be used to determine the distance up to a given redshift. The same oscillations produce the familiar dominant characteristics seen in the CMB power spectrum.

The BINGO (\textbf{B}AO from \textbf{I}ntegrated \textbf{N}eutral \textbf{G}as \textbf{O}bservations) telescope is a proposed new instrument designed specifically for observing such a signal and to provide a new insight into the Universe at $ z \lesssim 0.5$ with a dedicated instrument. The optical project consists of a compact, two 40-m diameter, static dishes with an exceptionally wide field-of-view (15 deg.) and very sensitive optics. The chosen technique for BAO observations is known as Intensity Mapping (e.g., \cite{Peterson2006}). The main scientific goal for BINGO is to claim the first BAO detection in the radio band, mapping the 3D distribution of HI, yielding a fundamental contribution to the study of dark energy, with observations spanning several years. 

The BINGO consortium is composed by Universidade de S\~ao Paulo and Instituto Nacional de Pesquisas Espaciais (Brazil), University of Manchester and University College London (United Kingdom), ETH Z\"urich (Switzerland) and Universidad de La Rep\'ublica (Uruguay). An earlier version of the BINGO telescope concept is described in a paper by R. Battye and collaborators \cite{Battye2013} and a comprehensive analysis of foregrounds and 1/f noise that can limit the BINGO performance was published by A. Bigot-Sazy and collaborators \cite{Bigot-Sazy2015}.

The structure of this paper includes this Introduction, followed by a description of the science motivation in Section \ref{science}. Section \ref{instrument} will briefly describe BINGO design. The current status of the project will be summarized in Section \ref{status}, and will be followed by the Concluding Remarks, in Section \ref{closing}

\section{The science}
\label{science}

Constraints on various existing dark energy models (see, e.g., \cite{Wang2005, Wang2007, Feng2008, Micheletti2009, He2009, Abdalla2010, He2011}) coming from BAO, such as those performed by Eisenstein and collaborators and Costa and collaborators \cite{Eisenstein2005, Anderson2014} since BAOs provide precision distance measurements up to high redshifts ($z \gtrsim 1$). These will allow the degeneracies between cosmological parameters inferred from the cosmic microwave background to be broken, constraining most extensions to the standard cosmological model which are usually degenerate with the Hubble constant. 

Measurements of the distribution of galaxies using observations in the radio region of the electromagnetic spectrum are based on the physics of the $21~cm$ (HI) line, which is a fundamental product of neutral hydrogen. HI redshift data allow us to construct a unique three­dimensional map of the mass distribution, providing a different view when compared to the maps obtained with optical telescopes. Indeed, HI is expected to be one of the best tracers of the total (dark+baryonic) matter (see, e.g., \cite{Padmanabhan2015}). 

\begin{figure}[h]
\begin{center}
\includegraphics[width=10cm]{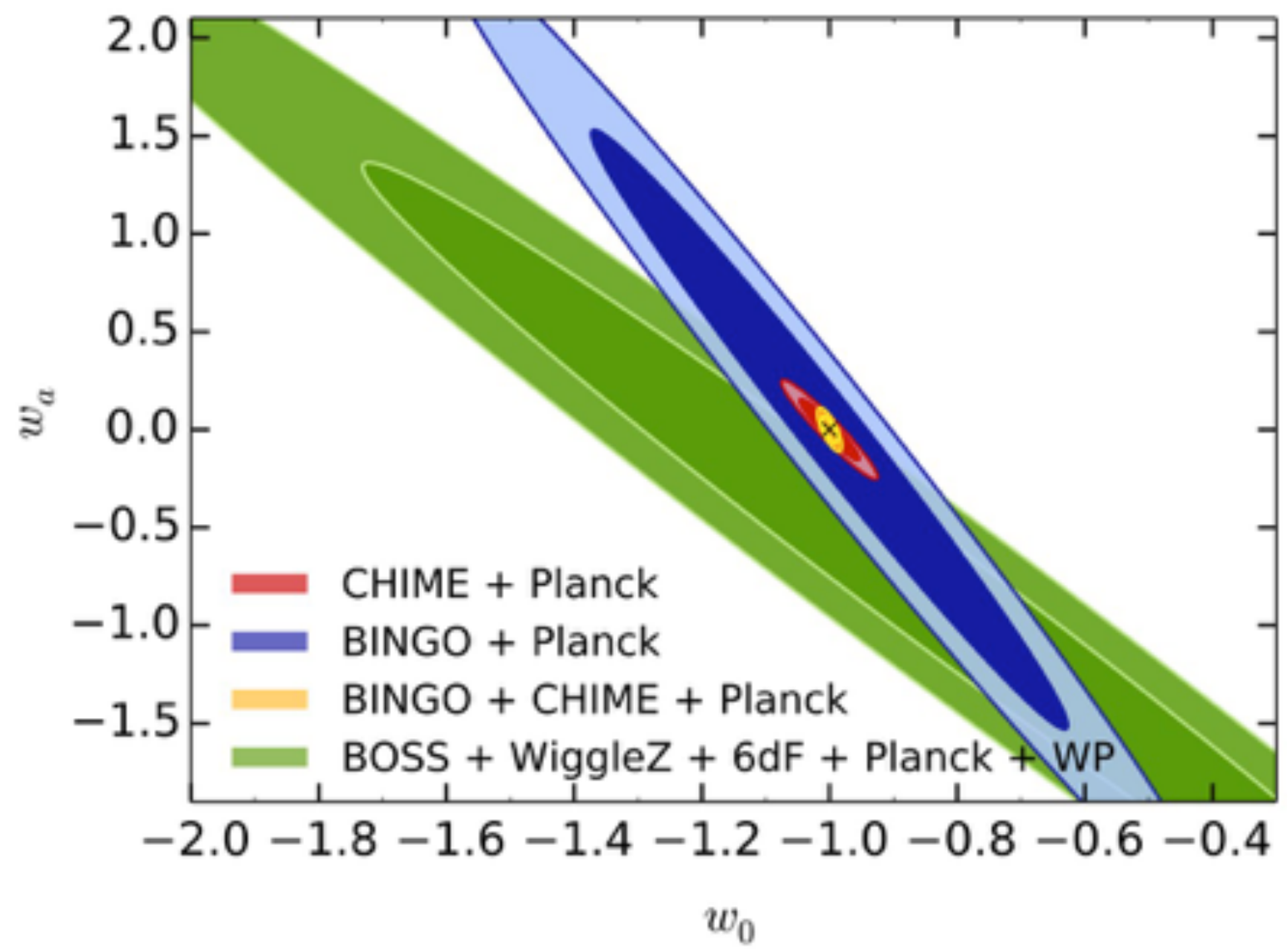}
\caption{Constraints on Dark Energy equation of state, including forecasts from CHIME and BINGO. Note the tight error elipse obtained with combined radio data (BAO from CHIME and BINGO + Planck).}
\label{BAOconstraints}
\end{center}
\end{figure}

BAO measurements can be used to calculate the likelihoods for cosmological parameters given different cosmological input data. Fig. \ref{BAOconstraints} shows the joint constraints for the equation-of-state of the dark energy with $w_{0}$ and $w_a$ ($1^{st}$ time derivative of $w_{0}$) given for various data sets. The constraints were computed using the Fisher Matrix code described in \cite{Bull2014}. It clearly shows the improvement obtained with the combination of the intensity mapping experiments BINGO and CHIME \cite{Newburgh2014}, compared to the current constraints given by the combination of the results of the CMB experiments Planck \cite{Ade2014, Ade2016} and WMAP (polarizations) \cite{Bennett2013}, with the optical galaxy surveys (BOSS \cite{Anderson2013, Tojeiro2014}, WiggleZ \cite{Kazin2014} and 6dF \cite{Beutler2011}. 

Furthermore, tight constraints can be made with radio data alone (BAO plus Planck), providing a completely independent measurement from optical surveys. The improvement in going from the red to yellow contours is due to having two independent measurements at different redshifts, showing the importance of BINGO. We have made estimates of the projected errors of the dark energy parameters assuming that all the other cosmological parameters are fixed. For constant $w$, the measurement would lead to an accuracy on $w$ of $\lesssim 8\%$ with BINGO experiment alone, including different types of foregrounds (\cite{Olivari2017}).  BINGO will be competitive with and complementary to optical surveys, which will likely be limited by different systematic errors.

\subsection{HI intensity mapping}

The standard approach to probing Large Scale Structure ($LSS$) is to perform a large redshift survey, to measure positions and redshifts of a large number of galaxies and use them to infer their density contrasts. Essentially, galaxies are being used as tracers of the underlying total matter distribution, a typical approach at optical wavelengths. The natural tracer at radio wavelengths is the $21~cm$ line of neutral hydrogen, but the volume emissivity associated with this line is low, meaning that detecting individual galaxies at $z \thicksim 1$ requires a very substantial collecting area. 

A number of approaches have been proposed to conduct intensity mapping surveys using an interferometer array rather than a single dish (see, for instance, \cite{Baker2011, Pober2012,Bemmel2012}. This approach can have a number of advantages, but it also requires complicated, and hence expensive, electronics to make the correlations. Using a single moderate sized telescope with an ultra stable receiver system is the lowest cost approach to intensity mapping measurements of BAO at $z~\lesssim 0.5$ (see, e.g., \cite{Battye2013}. BINGO will be the first telescope in the world operating in its frequency range whose goal is to study BAO with $21~cm$ intensity mapping. Its innovative idea is to use a telescope with broad beam operating at low frequencies to carry out intensity mapping \cite{Peterson2006, Loeb2008, Masui2013, Battye2013} and hence measure the overall integrated HI brightness temperature of a very large number of galaxies, used together as a LSS tracer. 

The difficulty in doing this is that the HI signal is typically $\thicksim 100~\mu K$ whereas the foreground continuum emission from the Galaxy is $\thicksim 1~K$ with spatial fluctuations $\thicksim 100~mK$. Fortunately, the integrated $21~cm$ emission exhibits characteristic variations as a function of frequency whereas the continuum emission has a very smooth spectrum. While there is a clear-cut statistical signature that allows for the two signals to be separated, the instrument used to distinguish the HI contribution from the foregrounds will need to be carefully designed – in particular to avoid systematic effects that can result in leakage of the continuum background into the HI signal. 

\section{The instrument}
\label{instrument}

Detecting signals of $\thicksim 100~\mu K$ with a receiver of standard performance implies that every pixel in our intensity map requires an accumulated integration time larger than 1 day over the course of the observing campaign, which is expected to last at least 3 -- 4 years. The total integration time can be built up by many returns to the same patch of sky but between these returns the receiver gains need to be highly alike and achieving this stability is a major design concern. The relative strength of the foreground, which is partially linearly polarized and concentrated towards the Galactic plane, means that the observations need to be made with a clean beam with low sidelobe levels and very good polarization purity. The general concept for the BINGO instrument is described in \cite{Battye2013} and updated in \cite{Battye2016}. 

A declination strip of $\thicksim 15^{\circ}$, centered at $\delta \thicksim -15^{\circ}$ aims at minimizing the Galactic foreground contamination and will be the optimal choice for the BINGO survey. The need to clearly resolve structures of angular sizes corresponding to a linear scale of around $150~Mpc$ at BINGO's chosen redshift range implies that the required angular resolution has to be $\thicksim 0^{\circ}.75$. 

\subsection{The optics}

\begin{figure}[h]
\begin{center}
\includegraphics[width=12cm]{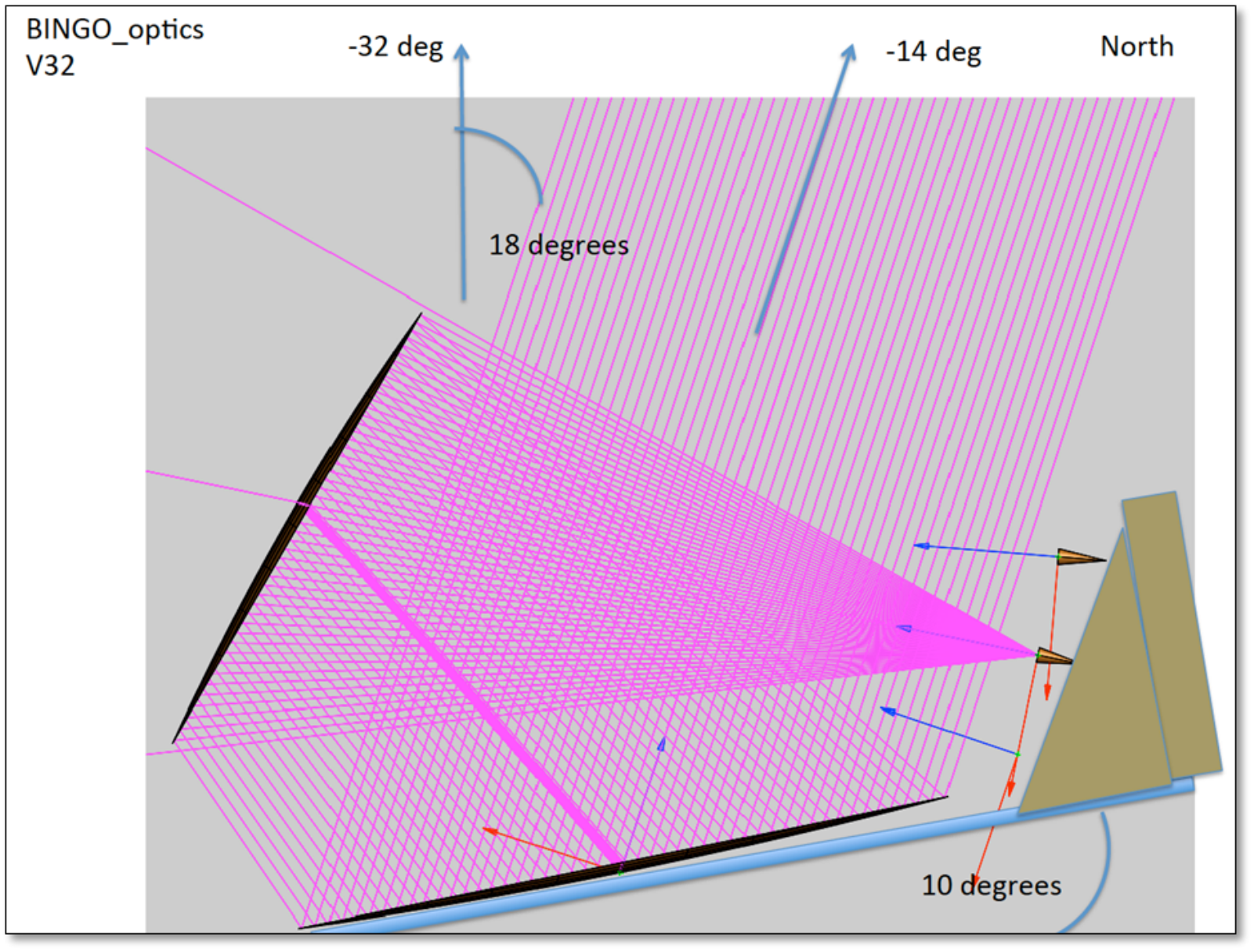}
\caption{The current optical design for BINGO}
\label{optics}
\end{center}
\end{figure}

The current optical design consists of two compact parabolic mirrors, with $\thicksim 40$ m diameter, which gives a wider field-of-view ($15^{\circ}$), lower sidelobe levels and improved polarization performance. Such a configuration has been used in CMB experiments, where there are similar strict requirements for low spillover and superb polarization performance. The configuration considers a primary fixed at $\thicksim 10^{\circ}$ in relationship to the ground, while the second will hang from a vertical structure, with a slightly negative inclination (see Fig. \ref{optics}). Since the mean wavelength of operation is $0.3$ m, the surface profile of the telescope mirrors should have an $rms$ error $\lesssim~15$ mm to achieve maximum efficiency. The telescope optics was designed by Dr. Bruno Maffei, also responsible for the design of the Planck HFI instrument feed horn antennas and their coupling to the telescope. 

The gain of the telescope beams associated with the feedhorns at the edge of the array are less than 1 dB lower compared with those from the feedhorns at the centre; optical aberrations are slight, the edge beams being almost circular. The number of feedhorns in the array could in principle be more than 100, but the project will use 50 horns, completely covering a $15^{\circ}$ strip of the sky every 24 hours

The guiding principle in the design of BINGO is simplicity. All components should be as simple as possible to minimize costs. Moreover, since there will be no moving parts, design, operation and instrument modeling will also be simpler than doing the same tasks for a conventional telescope. Another key advantage of a simple design is that it can be built quickly, allowing for results within a competitive science time window.  

\subsection{Receivers and feedhorns}

\begin{figure}
\begin{center}
\includegraphics[width=10cm]{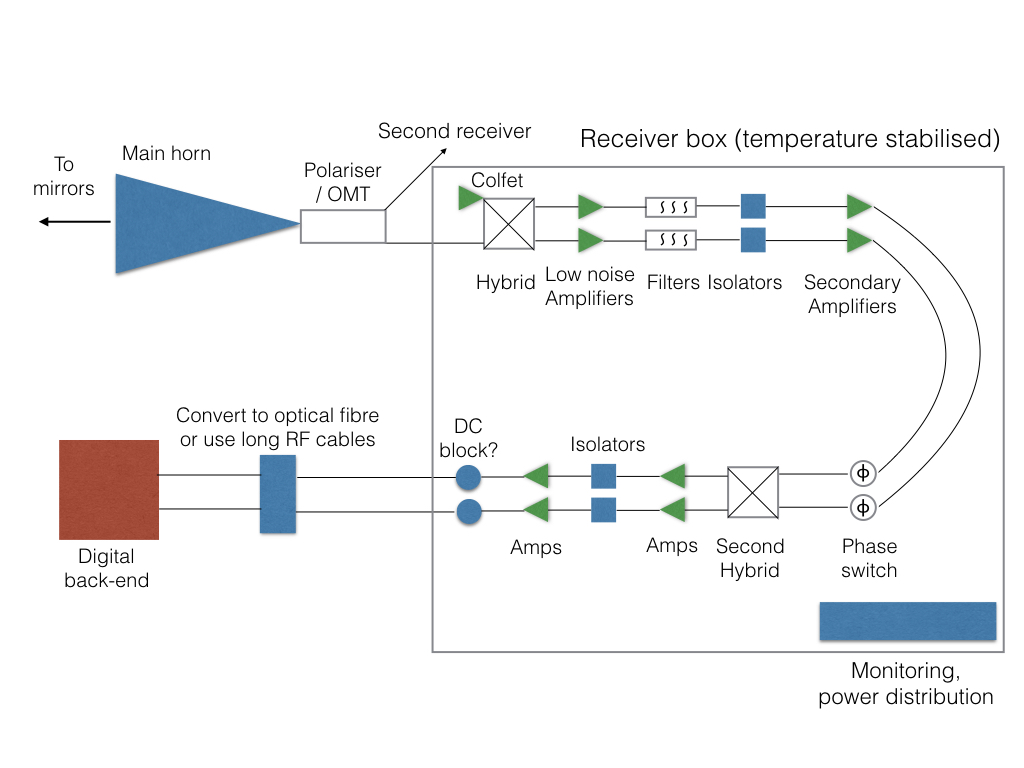}
\caption{A block diagram of the main components of the BINGO correlation receiver. The correlation is achieved by means of the hybrids, which in our case will be waveguide magic tees. It is expected the low noise amplifiers will achieve a system temperature of $\lesssim 50~K$.}
\label{receiver1}
\end{center}
\end{figure}

Each receiver chain contains a correlation system, as shown in fig. \ref{receiver1}, operating with uncooled amplifiers. The ambient temperature receivers will have a total system noise temperature of $\lesssim 50$ K and will operate in the frequency range $0.96 - 1.26$ GHz. This corresponds to a redshift range of $z = 0.13 - 0.48$, which is long after the reionization era. After a year of observation BINGO should achieve a one degree pixel noise of $60~\mu$ K in a single 1 MHz frequency channel. The current plan is to use uncooled amplifiers rather than cryogenically cooled ones because of their relatively good performance at these low frequencies and much lower cost. The sky background and spillover will contribute up to $10$ K and thus a cryogenically cooled receiver would give only a modest improvement in noise but with much more cost and complication. An additional benefit of receivers operating at ambient temperature is low maintenance given that cryogenics require constant attention that would be expensive to provide in a remote site.

\begin{figure}[h]

\begin{center}
\begin{minipage}{7.5cm}
\includegraphics[width=7cm]{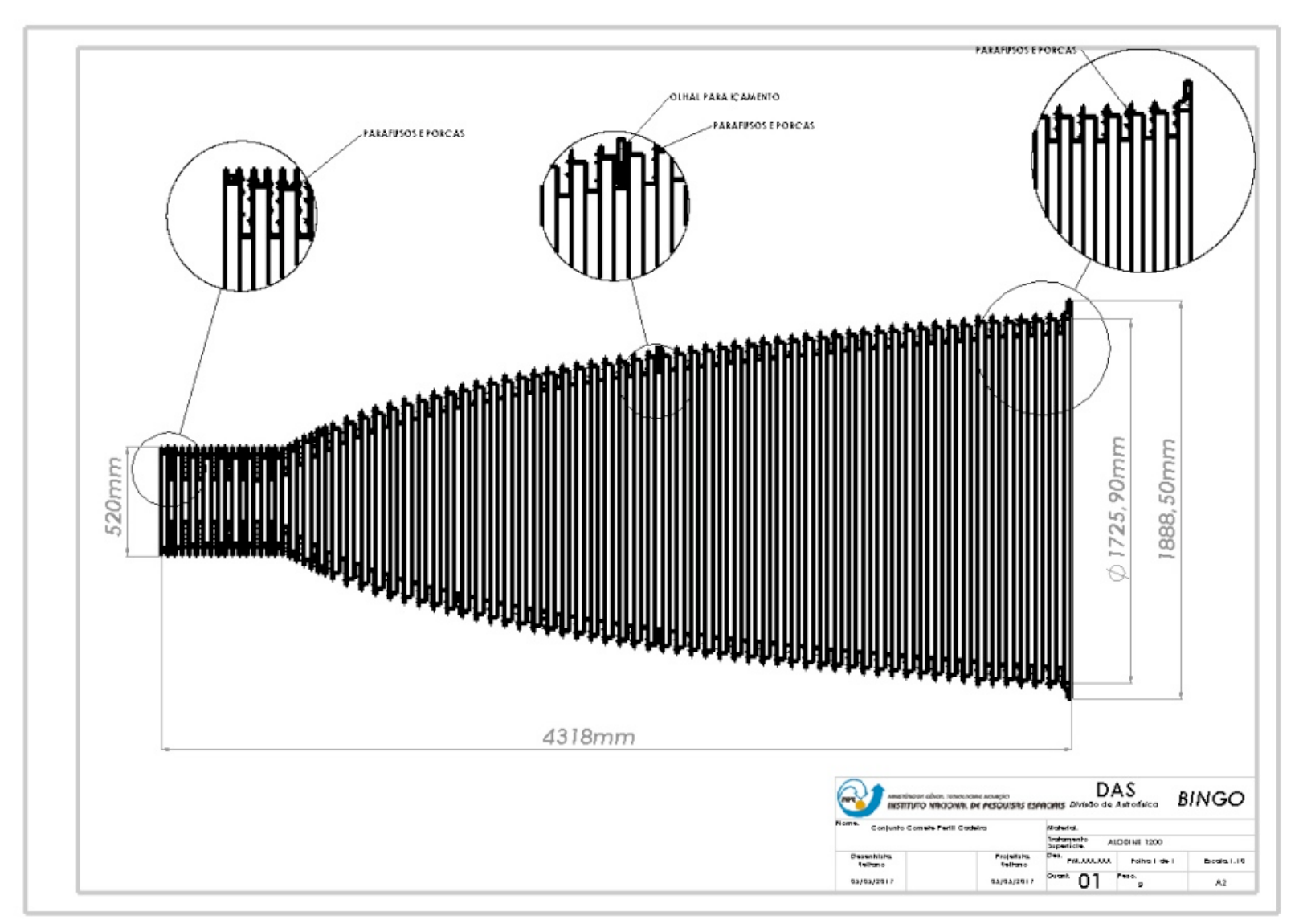}
\end{minipage}
\begin{minipage}{7.5cm}
\includegraphics[width=7cm]{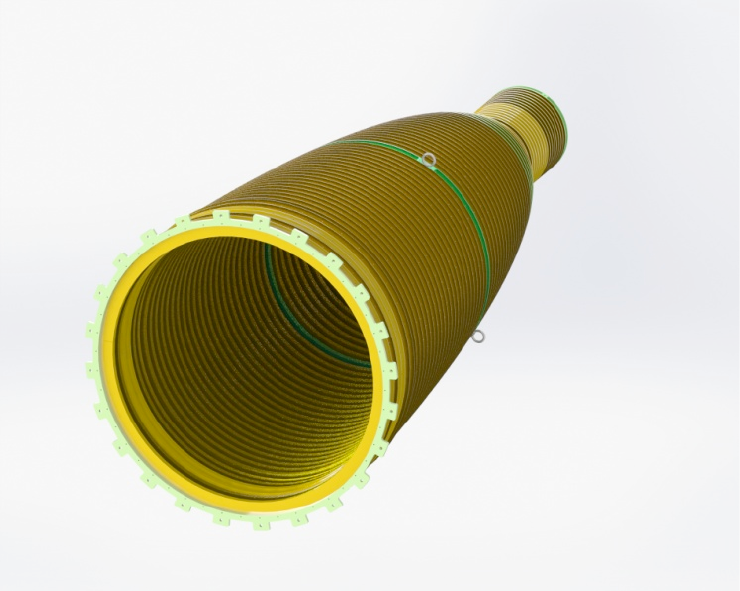}
\end{minipage}
\caption{Left: Horn blueprint, with details on the corrugation. Right: Horn mouth. Green rings are machined and contain sockets for suspension rings.}
\label{horns1}

\end{center}
\end{figure}

\begin{figure}[h]

\begin{center}
\begin{minipage}{7.5cm}
\includegraphics[width=7cm]{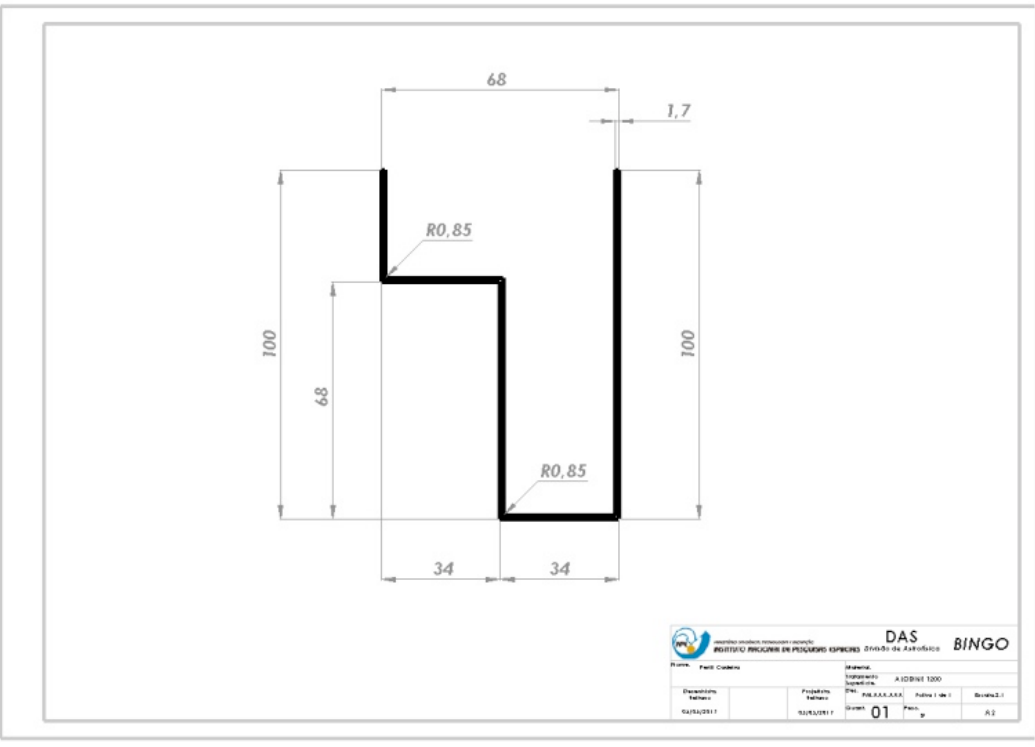}
\end{minipage}
\begin{minipage}{7.5cm}
\includegraphics[width=7cm]{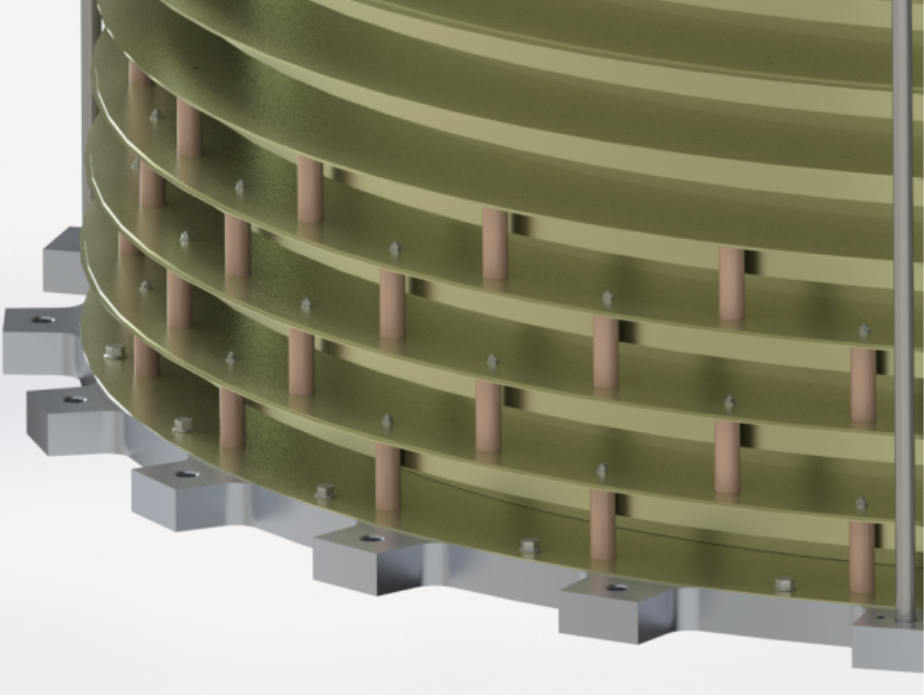}
\end{minipage}
\caption{Left: Profile used to produce each section of the horn, with corresponding dimensions. Right: Details of the section assembly close to the mouth. Note the machined gray section at the bottom which works as one of the structural parts of the horn.}
\label{horns2}

\end{center}
\end{figure}

BINGO will use specially designed conical corrugated feedhorns to illuminate the secondary mirror of the telescope. These need to be corrugated in order to provide the required low sidelobes coupled with very good polarization performance. Because of the large focal ratio needed to provide the required wide field of view, the feedhorns must be large with $\thicksim 1.7$ m in diameter and $\thicksim 4.9$ m in length (fig~\ref{horns1}). The electromagnetic design of such feedhorns is well understood; the challenge is to manufacture them at low cost and to minimize their weight, due to the uncommon size and number of horns needed. The mechanical project was done in Brazil, following the optical design by Bruno Maffei. It consists of ``4-like'' profiles which will be curved into 127 circular sections, assembled together to make the horns (fig.~\ref{horns2}). The final structure will  receive an alodyne coating to avoid weather deterioration. Also, it will receive an external layer of foam to ensure a better thermal stability during operation. A prototype is currently under construction and should be delivered by late October 2017. 

\subsection{Site selection}

The initial requirements for the telescope site were the following: a) a low Radio Frequency Interference (RFI) environment, distant for large cities or windmills; b) a topography that can easily support the two mirror design; c) easy access for researchers and technical staff;  d) no planning/environmental restrictions on building a telescope structure; and e) a clear view of the South Galactic Pole for the reference feedhorns, in practice meaning a latitude south of $\thicksim 30^{\circ}$. Recently, requirement e) was abandoned mostly due to budget constraints. 

\begin{figure}[ht]
\begin{minipage}{7.5cm}
\includegraphics[width=7cm]{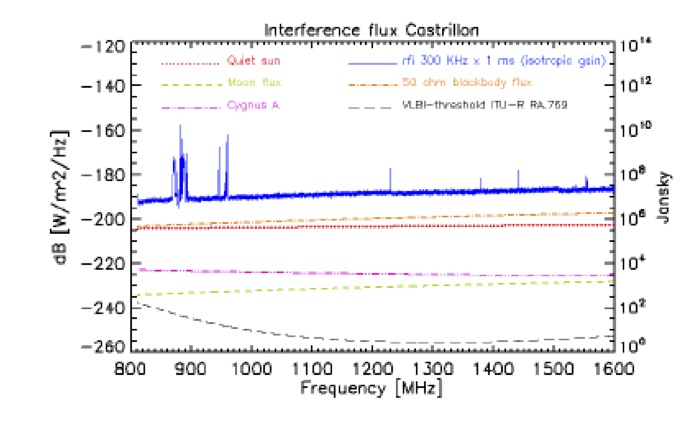}
\end{minipage}
\begin{minipage}{7.5cm}
\includegraphics[width=7cm]{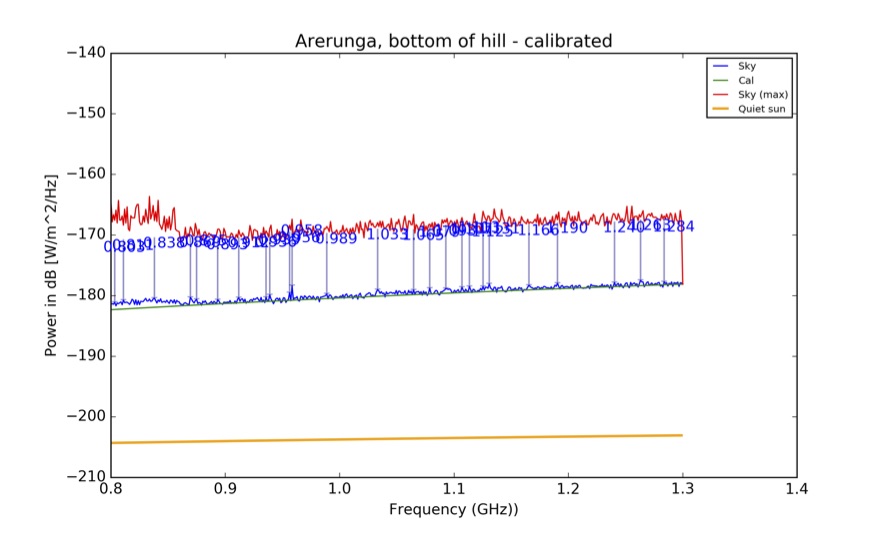}
\end{minipage}
\caption{RFI calibrated measurements in Castrillon (left) and Arerungu\'a (right). Note the cell phone peaks in the Castrillon measurements, around 950 MHz, and the absence of any clear signal, except for a small unknown excess power at lower frequencies in the Arerungu\'a measurement.}
\label{RFI1}
\end{figure}

The initial site was an abandoned gold mine in Castrillon, near the city of Minas de Corrales (Uruguay), which was discarded in 2016 due to problems with land ownership. A military farm, in Arerunguá, was indicated as a substitute in late 2016. The Arerungu\'a site had a very good RFI environment, but budget concerns led to another site selection campaign in Brazilian areas. A sample of RFI measurements taken in Castrillon and  Arerungu\'a is shown in Fig.~\ref{RFI1}.

\begin{figure}[ht]
\begin{minipage}{7.5cm}
\includegraphics[width=7cm]{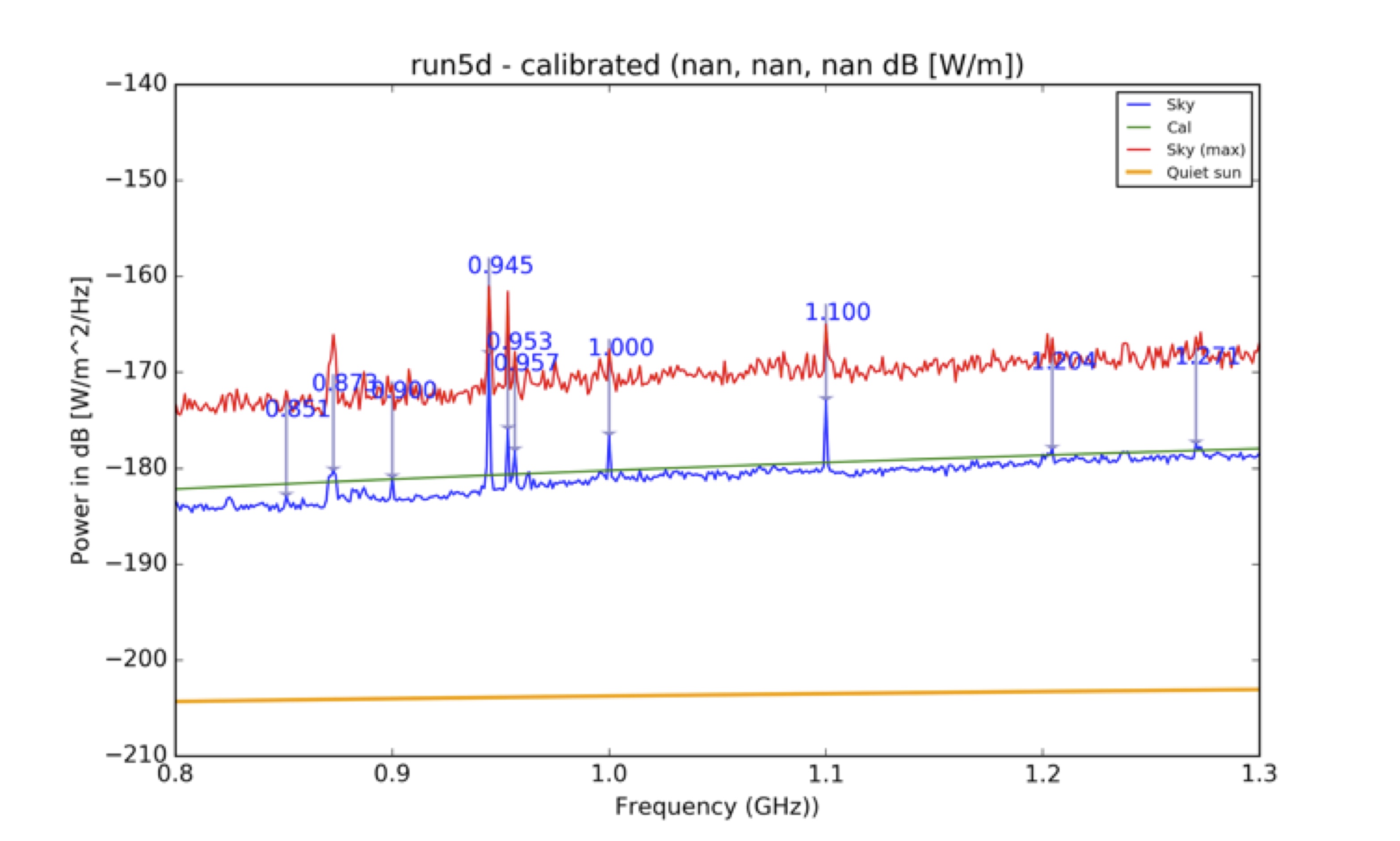}
\end{minipage}
\begin{minipage}{7.5cm}
\includegraphics[width=7cm]{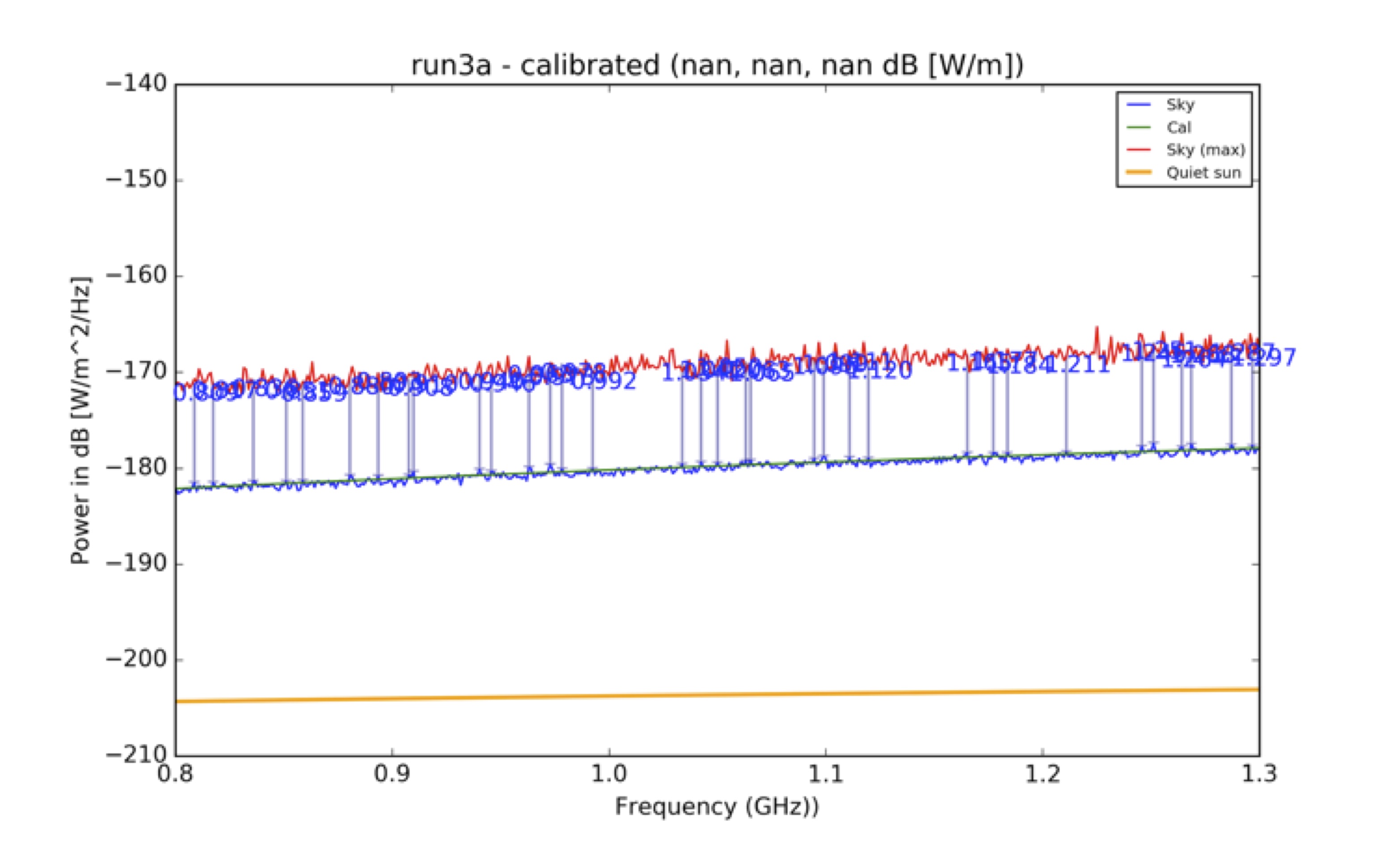}
\end{minipage}
\caption{RFI calibrated measurements in V\~ao do Gato (left) and Serra do Urub\'u (right). Note the cell phone peaks in the V\~ao do Gato measurements, around 950 MHz, and the absence of any clear signal on the Serra do Urub\'u measurement.}
\label{RFI2}
\end{figure}

\begin{figure}[ht]
\begin{center}
\includegraphics[width=18cm]{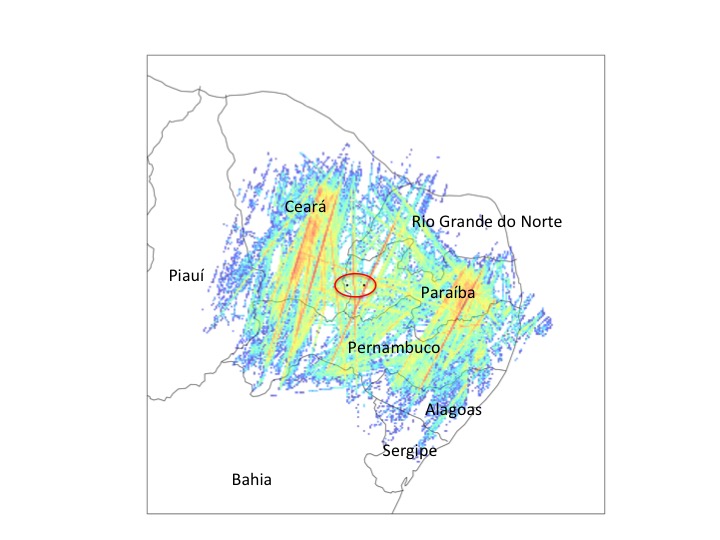}
\caption{Heat map of airplane routes covering the potential BINGO sites. Best sites are the two black dots in the center. The figure contains duplicate points. The data runs from August 22 to September 15, 2017.}
\label{airplane_routes}
\end{center}
\end{figure}

After a few months of procurement, including two regional centers of INPE, two sites were found in the west of Para\'{\i}ba, a state located in the northeastern part of Brazil. Besides being very clean in terms of RFI, the local topography and support from a federal university with several campii spread around the western area of the state, made them the preferable sites. The construction is planned to start in early 2018. RFI measurements taken in V\~ao do Gato and Serra do Urubú (both in Para\'{\i}ba) are shown in Fig~\ref{RFI2}. The airplane routes covering the state of Para\'{\i}ba, with V\~ao do Gato and Serra do Urub\'u sites marked are also shown in Fig.~\ref{airplane_routes}. 

\section{Current status}
\label{status}

\begin{figure}[h]
\begin{center}
\includegraphics[width=12cm, angle=90]{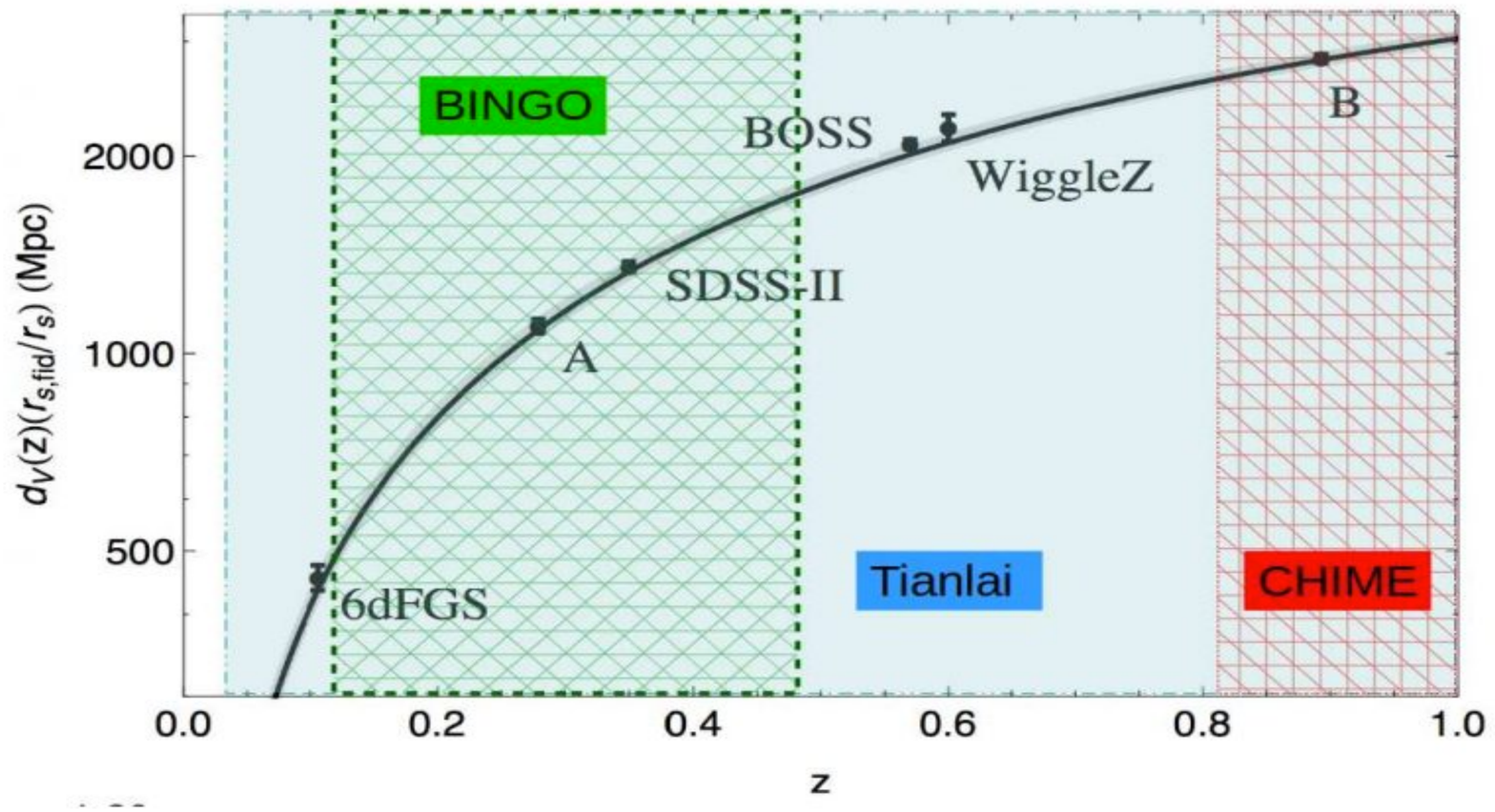}
\caption{A version of the Hubble diagram including BINGO, FAST and CHIME, as well as other optical surveys such as SDSS, 6dFGS, BOSS and WiggleZ. Note the complementary coverage in redshift between BINGO and CHIME. Tianlai (FAST) will encompass both surveys from BINGO and CHIME.}
\label{BINGO_others}
\end{center}
\end{figure}

BINGO systems are under planning or construction as of July 2017. Horns are being fabricated and prototypes should be tested at the Laboratory of Integration and Tests of satellites (LIT), at INPE, starting mid November 2017. Transitions, polarimeters, magic tees and the receiver are already designed; fabrication of receivers and transitions should start at the end of 2017, subject to details of importation of components. Testing of amplifiers and other receiver components has already started at INPE. 

The current available funding is circa US\$ 3.0 million, dedicated to the construction of the telescope. Total cost of BINGO is estimated to be around US\$ 4.2 million, the difference of US\$ 1.2 million being for site preparation and structure construction. Current BINGO's competitors are FAST (Five-hundred-meter Aperture Spherical Telescope) \cite{FAST2017} and CHIME (Canadian Hydrogen Intensity Mapping Experiment) \cite{CHIME2017}.

FAST main scientific goals involve hydrogen intensity mapping, fast radio bursts, pulsar timing, detection of interstellar molecules and interstellar communication signals. Midia references state that FAST will not be fully operational before late 2019, with three years needed to calibrate the various instruments. Once it is, it will likely require hundreds of radio astronomers. Problems with RFI produced by local tourists' visitation and also by its own actuators are current issues that may prevent FAST to operate at full capacity for a longer time.

Chime will also pursue the same major scientific goals as BINGO and FAST. Alike BINGO, it will have no moving parts but, instead of conventional mirrors and horns, it is based upon 20-m $\times$ 100-m long cylindrical mirrors and phased array beams. In addition to BAO and fast radio bursts, CHIME may also contribute to gravitational waves studies. It has entered operation in September 2017 and is able to monitor about 50\% of the sky per observing day.

Fig.~\ref{BINGO_others} shows the redshift space mapped by BINGO, FAST and CHIME, superimposed with current optical surveys.

\section{Final remarks}
\label{closing}
BINGO is an innovative project to measure BAOs in the redshift interval $0.12 \leq z \leq 0.48$, at radio wavelenghts. The measured signal is produced by the redshifted 21~cm line from neutral hydrogen, through a technique known as intensity mapping. BINGO will provide independent cosmological data and will probe the same redshift interval as the most important optical BAO surveys, but with different systematics. It is currently being constructed in Brazil, and will located in Para\'{\i}ba, Northeastern Brazil, with the local support of the Universidade Federal de Campina Grande. 

Scheduled to operate for at least 4 years, BINGO will provide high quality data, covering a wide range of scientific areas from Cosmology to Galactic science. With an upgrade to the digital backend, BINGO may also be able to detect a number of Fast Radio Bursts (FRBs) of which only $\sim$ three dozens have been currently detected so far and whose origin is unknown.

\section*{Acknowledgements}
CAW acknowledges the CNPq grant 313597/2014-6 and FAPESP Thematic project grant 2014/07885-0. 


\end{document}